\documentclass[12pt]{article}

\newcommand{\mrm}[1]{\mathrm{#1}}

\newcommand{\Al}{{\sc{Aleph}}}
\newcommand{\De}{{\sc{Delphi}}}
\newcommand{\Lt}{{\sc{L3}}}
\newcommand{\Op}{{\sc{Opal}}}
\newcommand{\Sl}{{\sc{Sld}}}
\newcommand{\Je}{{\sc{Jetset}}}
\newcommand{\Ar}{{\sc{Ariadne}}}
\newcommand{\He}{{\sc{Herwig}}}

\newcommand{\alphas}{\alpha_{\mrm{s}}}

\renewcommand{\c}{\mrm{c}}

\newcommand{\e}{\mrm{e}}

\newcommand{\g}{\mrm{g}}

\newcommand{\p}{\mrm{p}}
\newcommand{\q}{\mrm{q}}

\newcommand{\W}{\mrm{W}}

\newcommand{\cbar}{\overline{\mrm{c}}}

\newcommand{\pbar}{\overline{\mrm{p}}}
\newcommand{\qbar}{\overline{\mrm{q}}}

\newenvironment{Itemize}{\begin{list}{$\bullet$}%
{\setlength{\topsep}{0.2mm}\setlength{\partopsep}{0.2mm}%
\setlength{\itemsep}{0.2mm}\setlength{\parsep}{0.2mm}}}%
{\end{list}}
\newcounter{enumct}
\newenvironment{Enumerate}{\begin{list}{\arabic{enumct}.}%
{\usecounter{enumct}\setlength{\topsep}{0.2mm}%
\setlength{\partopsep}{0.2mm}\setlength{\itemsep}{0.2mm}%
\setlength{\parsep}{0.2mm}}}{\end{list}}

\newlength{\captivewidth}
\newlength{\abstwidth}
\def\ap#1#2#3   {{\em Ann. Phys. (NY)} {\bf#1} (#2) #3.}
\def\apj#1#2#3  {{\em Astrophys. J.} {\bf#1} (#2) #3.}
\def\apjl#1#2#3 {{\em Astrophys. J. Lett.} {\bf#1} (#2) #3.}
\def\app#1#2#3  {{\em Acta. Phys. Pol.} {\bf#1} (#2) #3.}
\def\ar#1#2#3   {{\em Ann. Rev. Nucl. Part. Sci.} {\bf#1} (#2) #3.}
\def\cpc#1#2#3  {{\em Computer Phys. Comm.} {\bf#1} (#2) #3.}
\def\err#1#2#3  {{\it Erratum} {\bf#1} (#2) #3.}
\def\ib#1#2#3   {{\it ibid.} {\bf#1} (#2) #3.}
\def\jmp#1#2#3  {{\em J. Math. Phys.} {\bf#1} (#2) #3.}
\def\ijmp#1#2#3 {{\em Int. J. Mod. Phys.} {\bf#1} (#2) #3.}
\def\jetp#1#2#3 {{\em JETP Lett.} {\bf#1} (#2) #3.}
\def\jpg#1#2#3  {{\em J. Phys. G.} {\bf#1} (#2) #3.}
\def\mpl#1#2#3  {{\em Mod. Phys. Lett.} {\bf#1} (#2) #3.}
\def\nat#1#2#3  {{\em Nature (London)} {\bf#1} (#2) #3.}
\def\nc#1#2#3   {{\em Nuovo Cim.} {\bf#1} (#2) #3.}
\def\nim#1#2#3  {{\em Nucl. Instr. Meth.} {\bf#1} (#2) #3.}
\def\np#1#2#3   {{\em Nucl. Phys.} {\bf#1} (#2) #3.}
\def\pcps#1#2#3 {{\em Proc. Cam. Phil. Soc.} {\bf#1} (#2) #3.}
\def\pl#1#2#3   {{\em Phys. Lett.} {\bf#1} (#2) #3.}
\def\prep#1#2#3 {{\em Phys. Rep.} {\bf#1} (#2) #3.}
\def\prev#1#2#3 {{\em Phys. Rev.} {\bf#1} (#2) #3.}
\def\prl#1#2#3  {{\em Phys. Rev. Lett.} {\bf#1} (#2) #3.}
\def\prs#1#2#3  {{\em Proc. Roy. Soc.} {\bf#1} (#2) #3.}
\def\ptp#1#2#3  {{\em Prog. Th. Phys.} {\bf#1} (#2) #3.}
\def\ps#1#2#3   {{\em Physica Scripta} {\bf#1} (#2) #3.}
\def\rmp#1#2#3  {{\em Rev. Mod. Phys.} {\bf#1} (#2) #3.}
\def\rpp#1#2#3  {{\em Rep. Prog. Phys.} {\bf#1} (#2) #3.}
\def\sjnp#1#2#3 {{\em Sov. J. Nucl. Phys.} {\bf#1} (#2) #3.}
\def\spj#1#2#3  {{\em Sov. Phys. JEPT} {\bf#1} (#2) #3.}
\def\spu#1#2#3  {{\em Sov. Phys.-Usp.} {\bf#1} (#2) #3.}
\def\zp#1#2#3   {{\em Zeit. Phys.} {\bf#1} (#2) #3.}

\begin{document}

\sloppy

\pagestyle{empty}

\begin{flushright}
LU TP 95--19 \\
hep-ph/9509235 \\
August 1995
\end{flushright}

\vspace{\fill}

\begin{center}
{\LARGE\bf QCD Physics Lessons of Z$^0$ %
Decays$^{\mbox{\normalsize *}}$}\\[10mm]
{\Large Torbj\"orn Sj\"ostrand} \\[2mm]
Department of Theoretical Physics, University of Lund, \\[1mm]
S\"olvegatan 14A, S-223 62 Lund, Sweden
\end{center}

\vspace{\fill}

\begin{center}
{\bf Abstract}\\[2ex]
\vspace{-0.5\baselineskip}
\noindent
\begin{minipage}{\abstwidth}
This talk contains a subjective selection of interesting results on
Z$^0$ decays, presented by the LEP and SLC groups. The emphasis
is on soft and semihard QCD physics. Results are put in a theoretical
context, and the limits of our current understanding are stressed.
Topics covered include event measures, prompt photons, coherence and
string effects, data and theory for particle rates and spectra,
particle correlations and Bose--Einstein effects.
\end{minipage}
\end{center}

\vspace{\fill}

\noindent
\rule[2mm]{50mm}{0.3mm}\\
$^{\mbox{\normalsize *}}$To appear (somewhat reduced to fit space
constraints) in the Proceedings of the International Europhysics
Conference on High Energy Physics, Brussels, Belgium,
July 27 -- August 2, 1995.\\

\clearpage
\pagestyle{plain}
\setcounter{page}{1}

\section{Introduction}

New QCD physics results continue to pour out from the \Al, \De, \Lt,
\Op\ and \Sl\ collaborations. Increased statistics, improved detectors
and refined data analyses make for more precise tests of physics
concepts. The task of the current minireview is not to cover everything
we know, but only to comment on some of the results of the year.
Other minireviews may be found in the
proceedings\cite{Ang,Fus,Met,Set,Mar,Wat,Ver}. Some attempts were made
to avoid overlaps --- for instance, hard QCD physics and $\alphas$
determinations are off limits --- but it has not always
been possible to define strict borders. Furthermore, the organizers
requested that emphasis be put on the physics lessons we have learned,
which means that I will take a more critical attitude to current
QCD-inspired models than is normally done in the experimental papers.
Therefore statements should not be accepted blindly, but hopefully they
may help stimulate a fruitful debate.

In our current standard picture of hadronic Z$^0$ events it is assumed
that the production process can be {\em factorized} into four steps:
\begin{Enumerate}
\item creation of the primary $\q\qbar$ pair (QFD);
\item additional parton production, described e.g. by parton showers
with angular ordering (perturbative QCD);
\item hadronization, described e.g. by string or cluster fragmentation
(nonperturbative QCD); and
\item secondary decays (QFD and nonperturbative QCD).
\end{Enumerate}
Even with such a simplified ansatz, the complexity makes
Monte Carlo event generators the preferred realization of physics
models.

\section{Event Measures and Prompt Photons}

Event shapes have been studied vigorously since the
{\sc{Petra}/\sc{Pep}} days, and extrapolations to Z$^0$
energies\cite{MK2,Phil} were presented long ago. Now when we sit with
the answer in hand it is impressive how well measures such as thrust
agree with predictions.

However, as emphasized by \De\cite{D548}, serious problems for models
are seen in properties related to out-of-the-plane
activity, such as aplanarity, four-jet rate and $p_{\perp\mrm{out}}$
spectrum. Since models normally only are matched to $O(\alphas)$
matrix elements, not to $O(\alphas^2)$ ones, some discrepancies could
have been expected and forgiven, but 30\% is uncomfortably large.

The problems are even more glaring when one turns to events with prompt
photons. Since $\q \to \q\gamma$ branchings compete with the
$\q \to \q\g$ ones, prompt photons probe the QCD evolution process.
In the past we have learned that the \Je\cite{TS} evolution, based on
emissions ordered in mass, is not doing well, while the
transverse-momentum-ordered \Ar\cite{LL} and the angularly-ordered
\He\cite{BW} do better though still not very well. Now new studies
are presented, e.g. by \Al\cite{A507} based on a democratic
jet finding algorithm, and further problems are found with models,
especially in the description of energetic photons.
It is less easy to find an excuse for these problems than for the
out-of-the-plane ones, so further model development appears mandatory.

\Lt\ has used prompt photons to study the properties of the
remainder-system\cite{L105}, to be compared with ordinary hadronic
events at a reduced energy. Whereas many properties vary in the
expected fashion, there are indications of a funny behaviour in
the soft-photon region. Here events ought to be fairly similar to
ordinary Z$^0$ decays without prompt photons; instead the data indicates
that jets are narrower on the average. Once the photon is sufficiently
energetic the expected behaviour is recovered, i.e. the scaled
jet width is larger the smaller the hadronic mass, in accordance with
the running of $\alphas$ and the character of nonperturbative
contributions. This could indicate yet another problem in the
description of photon emission, here for medium soft photons.

\section{Coherence and String Effects}

Interjet coherence is a perturbative phenomenon and the string effect
a nonperturbative one, but both are based on the same respect for the
colour flow topology of events, to leading order in $1/N_C^2$, where
$N_C=3$ is the number of colours.

\Op\cite{O332} compares $\q\qbar\gamma$ with $\q\qbar\g$ events.
In the angular
region between $\q$ and $\qbar$ the energy and particle flow is expected
to be higher in the former than in the latter event class. This is
indeed observed, and perturbative formulae\cite{coher} describe it
qualitatively. However, the separation is larger in the data than
expected by these formulae. Event generators describe the separation
fully when the colour flow is respected both in the perturbative and
the nonperturbative stages, but fail if either or both of these
requirements are relaxed.

\Al\cite{A518} instead compares the $\q\g$ and $\q\qbar$ angular regions
of $\q\qbar\g$ events. Again the former is expected to have a higher
energy and particle flow, and this is confirmed by the data.
In event generators that agree with data, about half the effect comes
from the perturbative stage and half from the nonperturbative one.
As above, models fail if colour flow effects are removed at either
stage.

\Al\ further demonstrates that ``jets are crooked'': if a quark jet
direction is reconstructed
in $\q\qbar\g$ events, then there is a tendency for high-momentum
particles to be found on the side away from the $\g$ jet and
low-momentum ones on the side towards the $\g$. The latter is what
could be expected from the soft emission by the colour dipole
spanned between the $\q$ and the $\g$; the former then follows as
a consequence of the jet direction being found as an average.

Also the particle--particle correlation and its asymmetry, defined
by analogy with the more familiar energy--energy correlation measures,
are good probes of colour topology effects. \Lt\cite{L116} and
\Al\cite{A455} agree that models are required to have both
perturbative and nonperturbative descriptions that respect this
topology.

In summary, we now see that a peaceful coexistence should be possible
between the perturbative coherence phenomenon and nonperturbative
colour-topology-based fragmentation models. Both are needed for a
complete description. The big loser may be
the local parton--hadron duality philosophy. According to LPHD the
energy and particle flows should be determined entirely by the
perturbative stage, with no extra structure added in the
nonperturbative stage. More about this in the next few paragraphs.

\section{Particle Rates and Spectra}

The collaborations have presented many new results on particle rates
and spectra. An extensive discussion on the data, including a
``world average'' table of particle rates per Z$^0$ event, is found
in the review by De~Angelis\cite{Ang}.

Let us begin by a discussion on the shape of spectra, leaving aside
absolute normalization. Generally these are well described by models.
They are also well described by the MLLA+LPHD framework, i.e.
modified leading-log approximation combined with local parton--hadron
duality\cite{MLLA}. This applies both to the shape of the distribution
and to the change of peak position (in the variable
$\xi = \log(1/x_p)$) as a function of c.m. energy.
A third prediction is that the peak position should vary in direct
relation to the mass of a particle. Here \Op\cite{O326} and
\De\cite{D539} compare a wide range of particles, from $\pi$ to
$\Xi$. The conclusion is unambiguous: peak positions do not agree
with the expected mass dependence. Good general agreement is found
with event generators, however, where secondary decays are major
sources of the observed particles. Only if such generators are used to
correct for the effects of secondary decays, i.e. if only primary
produced particles are considered, is the MLLA+LPHD pattern visible.

The LPHD assumption is not part of the QCD framework, unlike MLLA,
but an ad hoc ansatz that attempts to bypass the need for
hadronization models and treatment of particle decays. As such, it
has been given different interpretations by different authors, and in
its milder formulations it is a very useful guiding principle.
However, the hard-line approach, ``one parton, one hadron'', is now
shown to be in direct conflict with data on several counts.
It should be noted that programs such as \He\ or \Je\ are not
consistent with the hard-line LPHD school, not only on the issue
of secondary decays: in the programs the hadrons are not created at
the positions of partons but in the colour fields (clusters/strings)
spanned between partons. They agree with softer formulations of LPHD,
however, in that they give clear correlations between the local
phase-space density (and fluctuations) of partons and hadrons.

This does not mean that programs fully explain the shape of spectra.
For instance, the proton fraction of all charged particles drops at
large $x$, as demonstrated e.g. by SLD\cite{S205}, and this runs
counter to expectations. A possible explanation could be a suppression
of diquark production at small proper times of the fragmentation
process\cite{eden}.

We now turn to the average particle production rate per hadronic Z$^0$
event. Here much progress has been achieved in the last year, but the
picture still is not clear. For instance, many resonances are so broad
that the background subtraction is a very delicate process. Therefore
discrepancies may be found, for $\Delta^{++}$ between
$0.22\pm0.04\pm0.04$ by \Op\cite{O330} and $0.079\pm0.015$ by
\De\cite{D552}. Discrepancies such as these directly affect our
physics conclusions. One of the cleaner tests of production
dynamics is the relative rates of decuplet baryons, since
these should be little affected by resonance decays and isospin issues.
\Op\cite{O330} here finds a pattern in bad disagreement with models,
and also with a simple rule of a fixed suppression factor per extra
s quark:\\
\begin{tabular}{ccccc}
ratio & \Op\ & \Je\ & \He\ & ``world'' \\
$\Sigma^{*+}/\Delta^{++}$ & 0.086 & 0.206 & 0.373 & 0.177 \\
$\Xi^{*0}/\Delta^{++}$    & 0.029 & 0.029 & 0.089 & 0.049 \\
$\Omega^-/\Delta^{++}$    & 0.023 & 0.004 & 0.024 & 0.013 \\
\end{tabular}\\
If instead the ``world average'' numbers\cite{Ang} are used, agreement
with models improves significantly, and also a suppression factor of
about
0.2 per s quark is a good approximation. But even the ``world'' numbers
have experimental errors close to $\pm$50\%, so maybe the main
conclusion
is that it is too early to jump to conclusions.

Turning to the (non-)theory of particle composition, the production
rates could depend on several aspects, such as:
\begin{Enumerate}
\item flavour content of hadrons (e.g. K vs. $\pi$),
\item flavour compensation requirement (e.g. no baryon without
      an antibaryon),
\item spin (e.g. $\rho$ vs. $\pi$),
\item combined flavour+spin wavefunction (e.g. $\Sigma^0$ vs.
$\Lambda$),
\item mass (e.g. $\eta'$ vs. $\eta$),
\item phase space (e.g. cluster mass),
\item spatial shape of hadron wave function,
\item space--time overlap between adjacent wave functions,
\item topology of confinement field (e.g. g vs. q jets),
\item process type (e.g. $\e\p$ vs $\e^+\e^-$),
\item thermodynamics (e.g. in heavy-ion collisions), and
\item Bose--Einstein and other collective effects.
\end{Enumerate}
There is no basic law to forbid either of these dependences,
so at some level all of them should be expected. However, the hope
is that some are more important than others, so that it is possible
to find an efficient and simple first-order approximation.

Several approaches have been attempted; only a few are mentioned here.
\begin{Itemize}
\item The Lund\cite{Lund}/\Je\ approach contains a large number of
parameters
related to flavour and spin, and this number has tended to increase
over the years. One can more or less reproduce the data, but there is
very little predictive power left. Higher resonances (tensor mesons)
are more frequent in the data than expected; this could be reconciled
if these resonances would be polarized and tend to decay along the
global string direction, so that they could be viewed as intermediate
string pieces.
\item The UCLA model\cite{UCLA} is a variant of Lund in which the
string area law is taken to give the rate of flavour production as
well as the conventional kinematics dependence. It does a good job,
given the small number of parameters. Problems appear in the baryon
sector, however.
\item The \He\ cluster fragmentation approach also contains few
parameters and does well. One would therefore consider this a
very good first-order description, with the possibility of further
refinements, if needed. The cluster approach is faced with other
problems, however, as we will come to.
\item Chliapnikov and Uvarov\cite{Chl} have noted an interesting
regularity in production rates, consistent with an exponential
fall-off in $m^2$. This requires the introduction of ad hoc
isospin factors, however, and does not take into account the effects
of resonance feeddown, so it is difficult to draw any conclusions.
What is presented is also more of a fit than a physics scenario.
\item The most interesting new study is by Becattini\cite{Bec},
who takes a thermodynamical approach to the production rates.
Only three parameters are required: a temperature, a volume and an
s-quark suppression parameter. Within this constrained ansatz,
impressive agreement with the data is achieved. Hidden in the ansatz
is a number non-obvious assumptions, however, so the model is not fully
as constrained as it might seem. One also assumes that hadrons reach
complete thermal and chemical equilibrium, counter to conventional
wisdom that an $\e^+\e^-$ system is rapidly expanding and has a
rather low hadronic density. It will therefore be interesting to see
if new tests can be proposed to check the thermodynamics ideas.
\end{Itemize}
In summary, today we do not have one unique explanation of particle
production rates, but rather a set of mutually contradictory ideas.
This reflects our uncertainty about which are the main mechanisms at
play.

If there are some points in the list above that we would like
not to involve in an ultimate explanation, it would be 9 and 10,
since they break against such cherished notions as jet universality
and factorization. It is therefore notable that discrepancies now
are found in both areas. \Lt\cite{L092} has studied the
$\eta$ rate in three-jet events and observes a production in
excess of expectations in the lowest-energy jet, i.e. likely
the gluon jet. No anomaly is observed e.g. in the $\pi^0$ spectrum,
and other studies have shown that the shape of gluon jets is very
well predicted by models\cite{O315}, so this is a singular occurence.
It is too early to exclude a statistical fluctuation, but the
effect would be consistent e.g. with glueball production,
as allowed in some models\cite{Pet,Mon}. If so, even more spectacular
discrepancies could be expected for the $\eta'$.

There also appears to be problems with strangeness production, i.e.
mainly K and $\Lambda$. \De\cite{D540} notes a deficit of strange
particles in extreme two-jet events, i.e. events that
are not resolved into three or more jets even for small jet
resolution parameters.
Worse, both {\sc{Zeus}}\cite{Zeuss} and H1\cite{H1s} require
a s/u relative production rate of about 0.2, to be compared
with 0.3 in $\e^+\e^-$. Similar numbers have been presented since
many years, e.g. by neutrino experiments. Then it was at lower
energies, so, rightly or wrongly, problems in part could be blamed
on the poorly understood proton beam remnant. Now a low strangeness
production rate
is required in a broad range of rapidities and at large $p_{\perp}$.
Potentially this can be a devastating observation for our current
understanding of hadronization. Therefore this area should
be watched closely in the future, to see if we can find some clue
to what is going on.

\section{Correlations and Bose--Einstein Effects}

Once single-particle distributions begin to come under control,
it is interesting to turn to correlations for further information.

\Al\cite{A422} studies the angle between
the event axis and the $\p\pbar$ momenta in the rest frame of the
$\p\pbar$ pair. Data show that baryon production is preferentially
lined up along  the event axis, in agreement with the string model
but in sharp contrast to cluster models, where baryon-antibaryon
pairs are produced in isotropic cluster decays. An even more
precise test along these lines can be performed by \Sl\cite{S205}:
the large beam polarization implies that quarks preferentially
are found in one hemisphere and antiquarks in the other. The data
show that there are more energetic baryons than antibaryons in
quark jets, and the opposite in antiquark jets. Again this is in
in disagreement with the cluster picture. Also jet charge
studies\cite{A449} indicate that isotropic cluster decays give the
wrong flavour correlation pattern.

A hot topic at this meeting has been rapidity gaps at {\sc{Hera}}
and the Tevatron.
Conventional wisdom is that gaps should be very rare in $\e^+\e^-$
events, i.e. the rate should drop rapidly with increasing gap size.
Even current models for colour rearrangement\cite{Val,Jari} would
agree with this, but one could conceive of models with enhanced
rearrangement probability, which would then be disastrous for $\W$
mass determinations at LEP~2. Experimental studies fail to find
any unexpected effects\cite{A646,S210}.

The multiparticle production amplitude should be symmetrized
for identical bosons. This can lead to a Bose-Einstein (BE)
enhancement of particle production at small relative momentum
separation. Often this enhancement is parametrized as
$1 + \lambda \exp(-Q^2 R^2)$, where $\lambda$ is a chaoticity
parameter, $0 \leq \lambda \leq 1$, $R$ is the source radius
and $Q^2 = (p_1 - p_2)^2 = m_{12}^2 - 4m^2$. One could have
expected a source elongated along the event axis, but actually
$\e^+\e^-$ data are well described by the spherical ansatz.
The Gaussian shape is not in contradiction with data, but the
tests are not particularly discriminating. Typical values are
$R \approx 0.6$~fm and $\lambda \approx 1$, once effects of secondary
decays have been taken into account. Some studies come up with
$\lambda > 1$, which ought to be impossible. In a fitting
procedure the $\lambda$ and $R$ parameters are correlated, however,
so the evidence still is not fully compelling, in particular not if
also non-Gaussian shapes are considered. (One can observe
$\lambda \gg 1$ in $\pbar$n annihilation at rest, maybe related to
interference between specific channels\cite{Gas}. This is a warning
signal that not all of the low-$Q$ enhancement need be of BE origin.)

An experimental problem is that it is difficult to define a
reference sample without BE effects. On the theory side little
understanding exists of how exclusive event properties are
affected, e.g. if BE effects can modify the multiplicity of events.
A few explicit proposals have been made\cite{And,Lon}, but tests
are not yet sufficiently precise to discriminate.

Several new BE studies have appeared, see the review by
Verbeure\cite{Ver}. Especially nice is the first observation of
nontrival three-body correlations by \De\cite{D543}. The \Je\
BE approach partly describes these effects, but not fully.
Since the model was primarily based on two-particle correlations
this is maybe not surprising, but indicates the primitive state of
current modelling. An amazing aspect of this model is that it
implies a significant change in the $\pi^+ \pi^-$ mass spectrum,
with a downwards shift of the $\rho$ mass peak, as is now observed
in the data\cite{A509}. However, the way this effect arises in
the program is sufficiently subtle that one may worry whether
the BE explanation is not a red herring.

If BE effects these days attract increased attention, it is partly
due to the realization that BE correlations between the W$^+$ and
W$^-$ hadronic decay products at LEP~2 could affect the W mass
determination\cite{Lon}, just as colour rearrangement
could\cite{Val,Jari}.

\section{Summary}

The field of experimental QCD physics at the Z$^0$ is in impressively
healthy shape. It appears there are more new results than ever,
some improvements of previous studies but many breaking new ground.
Apologies to the authors of all those works that should have been
covered here but were not, for lack of space. Other interesting
results include the observation of $\g \to \c\cbar$ branchings at a
slightly high but still reasonable rate\cite{O291},
the first observation of $\Upsilon$ production\cite{D545,Oupsi},
the first attempts to measure a longitudinal polarization of the
$\Lambda$\cite{D707,O324},
the spin alignment of the phi\cite{O325},
and much more.

Most of the data is understood not only qualitatively but also
quantitatively. This is a success, not well reflected in this review.
However, it is natural to concentrate on the failures, which are the
ones that call out for more experimental and theoretical activity.

Possibly the most spectacular news is the potential breakdown of jet
universality and factorization, in the enhanced $\eta$ production rate
in gluon jets compared with quark jets, and especially in the reduced
strangeness production in $\e\p$ events compared with $\e^+\e^-$ ones.
These areas clearly need much further study, experimental and
theoretical.

The area of flavour production in general is in a bit of a crisis,
in that the models that served well for so many year now start to
show cracks under the onslaught of high-precision data. Maybe we need
a fresh outlook on the hadron production process.

Particle spectra still are reasonably well described by programs,
with some problems for baryons. However,
the hard-line interpretation of local parton--hadron duality
now is in conflict with data: it is not possible e.g. to neglect
the effects of unstable particle production and decay. In its more
flexible interpretations, LPHD still offers useful guidelines,
however.

The importance of the colour topology of events appears well
established,
and allows for a peaceful coexistence between perturbative and
nonperturbative manifestations.

Alignment along the colour flow axis is also visible in particle
correlations. Isotropically decaying clusters are excluded,
at least so long as clusters do not carry baryon number.

While the soft radiation pattern seems under control, problems may
be noted in the hard sector, both with respect to
out-of-the-event-plane properties and the emission of prompt photons.
It is too early to tell whether these are fundamental limitation
or are solvable technical problems.

Finally, it should be said that the need for QCD studies has not
decreased. On the contrary, there are many topics that deserve
continued attention in their own right, and for their importance to
other fields. As one such example, Bose-Einstein studies
not only may teach us about the particle production process but
also influence W mass determinations at LEP~2. There is therefore
every reason to\\
{\em{Keep up the good work!}}

\section*{Acknowledgments}
The impressive results obtained by the \Al, \De, \Lt, \Op\ and \Sl\
collaborations are what made this talk possible. The friendly help in
making results available is gratefully acknowledged. The blame for any
misinterpretation of the data rests with me.

\end{document}